\documentclass[a4paper,12pt]{article}
\pdfoutput=1
\usepackage{graphicx,rotating,colortbl,xcolor,wrapfig,hyperref,slashed}

\hypersetup{colorlinks,bookmarksopen,bookmarksnumbered,
linkcolor=blus,pdfstartview=FitH,urlcolor=rossos,citecolor=verde}

\def\to{\rightarrow}
\def\bea{\begin{eqnarray}}
\def\eea{\end{eqnarray}}

\def\mpl{M_{\rm Pl}}

\definecolor{Gray}{gray}{0.95}

\newcommand{\lh}{\hat{\lambda}}

\newcommand{\Rt}{\tilde{R}}

\definecolor{rosso}{cmyk}{0,1,1,0.4}
\definecolor{rossos}{cmyk}{0,1,1,0.55}
\definecolor{rossoc}{cmyk}{0,1,1,0.2}
\definecolor{blu}{cmyk}{1,1,0,0.3}
\definecolor{blus}{cmyk}{1,1,0,0.6}
\definecolor{bluc}{cmyk}{1,1,0,0.1}
\definecolor{verde}{cmyk}{0.92,0,0.59,0.25}
\definecolor{verdec}{cmyk}{0.92,0,0.59,0.15}
\definecolor{verdes}{cmyk}{0.92,0,0.59,0.4}

 \def\be   {\begin{equation}}   \def\ee   {\end{equation}}
 \def\ba   {\begin{array}}      \def\ea   {\end{array}}

\font\tenrsfs=rsfs10 at 12pt
\font\sevenrsfs=rsfs7
\font\fiversfs=rsfs5
\newfam\rsfsfam
\textfont\rsfsfam=\tenrsfs
\scriptfont\rsfsfam=\sevenrsfs
\scriptscriptfont\rsfsfam=\fiversfs
\def\mathscr#1{{\fam\rsfsfam\relax#1}}

\def\circa#1{\,\raise.3ex\hbox{$#1$\kern-.75em\lower1ex\hbox{$\sim$}}\,}

\newcommand{\beq}{\begin{equation}}
\newcommand{\eeq}{\end{equation}}

 \def\ex{\epsilon}
 
 \def\kx{\kappa}
 
 \def\lx{\lambda}

\def\ttau{{\tilde{\tau}}}

\def \lta {\mathrel{\vcenter
     {\hbox{$<$}\nointerlineskip\hbox{$\sim$}}}}
\def \gta {\mathrel{\vcenter
     {\hbox{$>$}\nointerlineskip\hbox{$\sim$}}}}

\begin{document}

\thispagestyle{empty}
\vspace{0.1cm}
\begin{center}
{\huge \bf \color{rossos} 
Black holes and Higgs stability}  \\[2cm]

{\bf\large Nikolaos Tetradis}  \\[5mm]

{\it Department of Physics, University of Athens, Zographou 157 84, Greece
\\
and 
\\
Physics Department, Theory Unit, CERN, CH-1211 Geneva 23, Switzerland}

\vspace{2cm}
{\large\bf\color{blus} Abstract}

\begin{quote}
We study the effect of primordial black holes on the classical
rate of nucleation
of AdS regions within the standard electroweak
vacuum. We find that the energy barrier for transitions to the 
new vacuum, which characterizes the exponential suppression 
of the nucleation rate,
can be reduced significantly in the black-hole background. 
A precise analysis is required in order to determine whether the the 
existence of primordial black 
holes is compatible with the form of the Higgs potential at high temperature 
or density in the Standard Model or its extensions. 
\end{quote}
\end{center}

\newpage
\tableofcontents
\newpage\normalsize


\section{Introduction}

The creation of an anti-de Sitter (AdS) bubble within asymptotically flat or 
de Sitter (dS) space is a problem relevant for the 
issue of vacuum decay. The nucleation of such a bubble can occur through
quantum fluctuations within the false vacuum, with a 
rate that is exponentially suppressed by the action of the 
saddle point dominating the transition \cite{deluccia}. The creation 
is also possible during inflation, when massless fields
fluctuate with a characteristic scale set 
by the almost constant Hubble parameter $H_{\rm inf}$. This process can be viewed as a transition beyond the potential barrier in the direction of the 
true vacuum, which is induced by the effective Gibbons-Hawking temperature $T=H_{\rm inf}/2\pi$~\cite{gibbons}. A similar transition may occur 
after the end of inflation, during reheating, or in any period during which
fluctuations become strong. 

Through the above processes 
the Higgs field of the Standard Model can fluctuate towards
values a few orders of magnitude below the Planck scale, for which its potential
becomes deeper than for the standard electroweak vacuum \cite{espinosa,instab,cosmo2,zurek,tetradis}.
Regions where the transition takes place can
be considered as regions of AdS space within standard Minkowski or dS space. 
If they are approximated as spherically symmetric,
their evolution becomes tractable analytically, along lines similar to
\cite{blau}. 
The general conclusion is that the AdS bubbles expand and engulf the 
whole external Minkowski space, evolving
into a singularity characterized as the AdS crunch \cite{deluccia,tetradis}. 
For bubbles created during inflation
this catastrophic scenario can be avoided
if $H_{\rm inf}$ is sufficiently small, so that fluctuations beyond the potential
barrier are highly improbable \cite{tetradis}.

It was pointed out recently \cite{gregory} that the 
quantum nucleation of bubbles of true vacuum can be enhanced around
black holes that act as impurities in the false vacuum.
In the context of the Standard Model, the presence 
of black holes may destabilize the standard electroweak vacuum
by reducing drastically the barrier for quantum fluctuations 
in the direction in which the Higgs potential becomes negative.
The origin of the primordial black holes that are relevant for this issue
lies in strong fluctuations in the early Universe, which affect the Higgs field
as well. 
A question that arises naturally is whether the  
creation of black holes
during inflation or during later periods is
accompagnied by the appearance of AdS bubbles around them that  
arise as classical fluctuations.
In such a scenario, the transition to the true vacuum is not a
quantum phenomenon, but is triggered by the high temperature or density
environment.

A primordial 
black hole can form when the density fluctuations are sufficiently large
for an overdense region to collapse \cite{pbh}. 
Its maximal mass is of order the total mass within the 
particle horizon $m_{\rm bh}\sim \mpl^2/H$,
while its maximal radius is $R_{\rm bh} \sim 1/H$.
These estimates are also valid for black holes that are pair-produced during
inflation \cite{hawking}. 
The presence of a significant number of primordial black holes
today depends on their production rate, their dilution by the
expansion and their evaporation rate. Black holes larger than roughly $10^{15}$
gr can survive until the present time and play the role of dark matter. 
The form of the
Higgs potential places an additional constraint on the 
possible existence of primordial black holes: The creation of 
an AdS bubble around a typical black hole must be disfavored. In the
opposite case, the Universe cannot evolve to its present form.

In the following we discuss the possible enhancement of classical instabilities 
of the standard electroweak vacuum near black holes. The relevant quantity,
which is the focus of our analysis, is the height of the energy barrier  
between the standard vacuum and the region in which the Higgs potential becomes 
negative.
The nucleation rate of the new phase 
is exponentially suppressed by the ratio of the energy barrier to the 
characteristic scale of fluctuations, such as the temperature or any
other parameter determining the field dispersion.

In section \ref{AdS} and appendices \ref{app1}-\ref{app3} we analyze 
the idealized problem of an AdS bubble with a black hole at its center,
within asymptotically flat or dS space. 
We determine the ADM mass of the critical bubble, 
from which the energy barrier can be estimated. 
In section \ref{central} we
study the effect of the black hole on the critical-bubble mass. 
In section \ref{estimate} we estimate the energy barrier using the energy scales
of the Higgs potential. In section \ref{numerical} we address the problem
by solving numerically the equation of motion of the Higgs field in order to
obtain the precise form of the bubble profile and the related energy barrier.
The last section contains our conclusions.

\section{Matching the geometries}\label{AdS}

An idealized setting in which our problem can be addressed 
consists of a spherical region of AdS-Schwarzschild space with a mass parameter
$m$, separated by 
a thin shell from a Schwarzschild or dS-Schwarzschild exterior with
a mass parameter $M$. The difference $\delta M=M-m$ determines how the 
mass $M$, 
attributed by a faraway observer to this configuration, compares to
the mass parameter $m$. The sign of $\delta M$ is a first indicator 
of whether the ``dressing" of a black hole by an AdS bubble is energetically 
favorable.

For an external observer, the presence of the bubble has a gravitational effect
equivalent to the presence of a central mass. As a result, the exterior metric 
is of the Schwarzschild or dS-Schwarzschild type.  
We assume a spherical AdS-Schwarzschild interior,
separated from the external space by a thin wall with constant surface tension $\sigma$.
In the thin-wall approximation,  the motion of the wall is determined by junction conditions that relate the extrinsic curvature
on each of its sides~\cite{israel}. The basic elements of the problem are:
\begin{enumerate}
\item  The space inside the bubble has a metric 
\be
 ds^2=-f_{\rm in}(r)\,  d\eta^2+ \frac{ dr^2}{f_{\rm in}(r)}+r^2  d\Omega^2_2,\qquad r< R,
\label{ads-metric} \ee
with $f_{\rm in}(r)=1+r^2/\ell_{\rm AdS}^2-{2Gm}/{r}$ and $G=1/(8\pi \mpl^2)$. 
The parameter 
$m$ characterizes the black hole located at $r=0$.
For $m=0$ we
obtain the standard AdS solution, expressed in global coordinates.
The space has negative vacuum energy $V_{\rm AdS}$,
corresponding to the length scale $1/\ell_{\rm AdS}^2=8\pi G |V_{\rm AdS}|/3$. 

\item The space outside the bubble is described by the metric 
\be
 ds^2=-f_{\rm out}(r)\,  dt^2+\frac{ dr^2}{f_{\rm out}(r)}+r^2  d\Omega^2_2,\qquad r >R,
\label{schw-metric} \ee
with $f_{\rm out}(r)=1-r^2/\ell_{\rm dS}^2-{2GM}/{r}$.
The positive vacuum energy $V_{\rm dS}$
is related to the length scale $\ell_{\rm dS}$ through $1/\ell_{\rm dS}^2=8\pi G V_{\rm dS}/3$. 
The parameter $M$ corresponds to the mass assigned to the bubble by an asymptotic observer.

\item
The two regions are separated by a domain wall with constant surface tension $\sigma$.
The metric on the domain wall can be written as
\be
 ds^2=- d\tau^2+R^2(\tau)  d\Omega^2_2,
\label{wall-metric} \ee
where $R(\tau)$ denotes the location of the wall in both coordinate systems (\ref{ads-metric}) and (\ref{schw-metric}).
The evolution is expressed in terms of the proper time $\tau$ on the wall. 
In the scenario of Higgs fluctuations produced during inflation, 
$\sigma$ is induced by the
kinetic and potential energy of the Higgs field.
\end{enumerate}

The continuity of the metric gives
\begin{eqnarray}
f_{\rm in}(R) \, \dot{\eta} &=& \ex_1\left(\dot{R}^2+f_{\rm in}(R) \right)^{1/2},
\label{doteta} \\
f_{\rm out}(R) \, \dot{t}&=&\ex_2\left(\dot{R}^2+f_{\rm out}(R) \right)^{1/2},
\label{dott} \end{eqnarray}
where $\ex_1=\pm 1$, $\ex_2=\pm 1$ are possible sign choices
and a dot denotes a derivative with respect to $\tau$.
We define 
a spacelike vector $\xi^\mu$ orthogonal to the 
four-velocity of the wall, given by 
$U^\mu_{\rm in}=(\dot{\eta},\dot{R},\vec{0})$ and 
$U^\mu_{\rm out}=(\dot{t},\dot{R},\vec{0})$
in each of the frames. 
We have
\begin{eqnarray}
\xi^\mu_{\rm in}&=&\left(\frac{\dot{R}}{f_{\rm in}},f_{\rm in} \, \dot{\eta},\vec{0}\right)
=\left(\frac{\dot{R}}{f_{\rm in}},\ex_1 (f_{\rm in}+\dot{R}^2)^{1/2},\vec{0}\right)
\label{perp1} \\
\xi^\mu_{\rm out}&=&\left(\frac{\dot{R}}{f_{\rm out}},f_{\rm out}\, \dot{t},\vec{0}\right)
=\left(\frac{\dot{R}}{f_{\rm out}},\ex_2(f_{\rm out}+\dot{R}^2)^{1/2},\vec{0}\right).
\label{perp2} \end{eqnarray}
The junction conditions connect the discontinuity in the extrinsic curvature to the surface tension \cite{israel}:
\be
(K_{\rm out})^i_j-(K_{\rm in})^i_{~j}=-4\pi G \sigma \delta^i_{~j}.
\label{junction} \ee
They give
\be
\ex_2(f_{\rm out}+\dot{R}^2)^{1/2}-\ex_1(f_{\rm in}+\dot{R}^2)^{1/2}=\beta_{\rm out}-\beta_{\rm in}=-4\pi G\sigma  R.
\label{junction12} \ee

The square of eq.~(\ref{junction12}) can be put in the form:
\be
2GM=2Gm+\left(\kappa^2-\frac{1}{\ell_{\rm AdS}^2}-\frac{1}{\ell_{\rm dS}^2} \right)R^3+2\ex_2 \kappa R^2
\left(1-\frac{2GM}{R}-\frac{R^2}{\ell_{\rm dS}^2}+\dot{R}^2 \right)^{1/2},
\label{square1} \ee
with $\kappa=4\pi G\sigma$.
For fixed $M$ and large $R$ 
the wall velocity is directly related to the volume contribution 
proportional to $\kappa^2-1/\ell_{\rm AdS}^2-1/\ell_{\rm dS}^2$. 
The total volume effect can be negative or positive.
As a result, it is possible for the total mass $M$ to be negative.
Alternatively, we can solve eq.~(\ref{junction12}) for $M$, with the
result
\be
2GM=2Gm-\left(\frac{1}{\ell_{\rm AdS}^2}+ \frac{1}{\ell_{\rm dS}^2}+\kappa^2 \right)R^3+2\ex_1 \kappa R^2
\left(1-\frac{2G m}{R}+\frac{R^2}{\ell_{\rm AdS}^2}+\dot{R}^2 \right)^{1/2}.
\label{square} \ee
For $\ex_1=1$,  $\dot{R}\ll1$ and $G\to 0$, the above expression has a Newtonian interpretation: 
The mass $M$ attributed to the bubble of AdS 
by an outside observer contains a volume term 
proportional to $-(1/\ell_{\rm AdS}^2+ {1}/{\ell_{\rm dS}^2}+\kappa^2)$. The contribution $-(1/\ell_{\rm AdS}^2+1/\ell_{\rm dS}^2)$ corresponds to the 
difference in vacuum energy density between the
interior and exterior of the bubble, while $-\kappa^2$ reproduces correctly 
the gravitational self-energy of the wall. The expansion of the 
second term in eq.~(\ref{square})
produces the surface energy of the bubble, the interaction energy between the 
surface and the black hole of mass $m$,
the surface-volume interaction energy 
and the kinetic energy of the wall \cite{blau}.

The evolution of a bubble can be deduced through the solution of 
eq. (\ref{junction12}).
By squaring this equation twice,
we can express the equation of motion for the bubble wall
as the equation
for the one-dimensional motion of a particle of constant `energy' 
in an effective `potential'. 
For fixed values of $\ell_{\rm AdS}$, $\ell_{\rm dS}$, $\kappa$, $m$, the
`energy' depends on the total mass $M$ of the configuration. A multitude of
wall trajectories are possible for various values of $M$, 
describing shrinking or expanding bubbles 
\cite{blau,tetradis}. 
A complete analysis is provided in appendices \ref{app1}-\ref{app3}. 
In general, for a given set of parameters there is a critical
configuration that separates small bubbles that tend to collapse from
large bubbles that tend to grow and engulf the external space. The mass $M$
of these bubbles characterizes the energy barrier for transitions towards the
deeper AdS vacuum. In the following section we focus on the dependence of
the energy barrier on the mass parameter $m$ of the central black  hole.

\section{The critical bubbles}\label{central}

We concentrate on the bubble evolution for $\kappa^2 \ll 1/\ell_{\rm AdS}^2$.
For vacuum energies and surface tension induced by 
the Higgs field $h$ through its potential, 
$\kx^2\sim (G\sigma)^2$ is suppressed by $h^2/\mpl^2$ relative to 
$1/\ell_{\rm AdS}^2$. As a result, the scenario with $\kx^2>1/\ell_{\rm AdS}^2$ can be realized 
only for values $h\gta\mpl$. Gravitational corrections
to the potential become important at such scales, so that it is not possible
to make firm predictions. Alternatively, one may assume that the Higgs field
varies substantially 
over sub-Planckian distances. Such a scenario lacks predictivity also, as higher-derivative terms in the effective action 
become dominant. For these reasons, we focus our analysis on the
case $\kx^2 \ll 1/\ell_{\rm AdS}^2$. 

%
\begin{figure}
\includegraphics[width=0.7\textwidth]{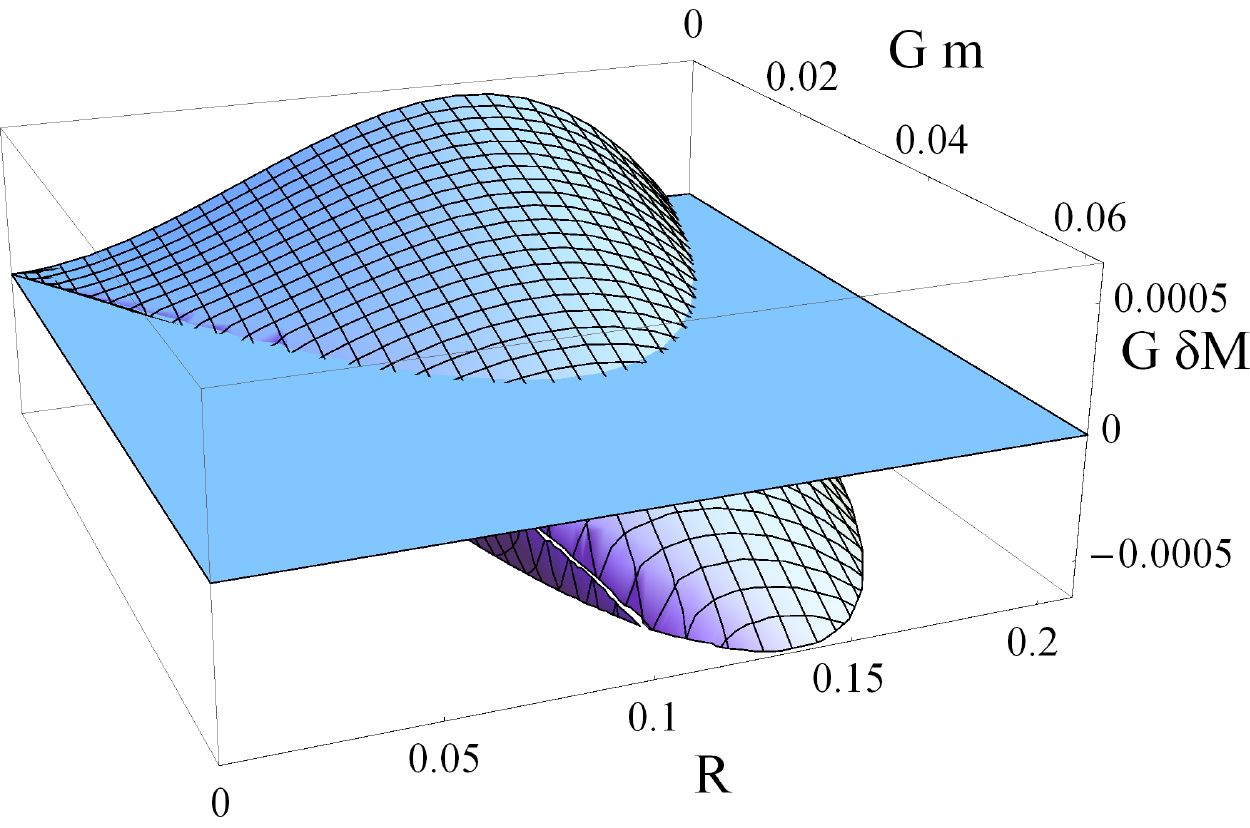}
\caption{The mass difference $\delta M=M-m$ for a bubble with $\dot{R}=0$ and $\kx=0.1$, $1/\ell_{\rm dS}\to 0$, 
as function of its radius $R$ and the mass $m$ of the central black hole. All quantities are given in units of $\ell_{\rm AdS}$.}
\label{fig1}
\end{figure}

It is reasonable to assume that at the time of production of an
AdS bubble the wall has small velocity.
The reason is that the walls of the random field fluctuations 
do not have a preferred direction of motion. 
In the context of the spherical approximation, the
assumption closest to the realistic scenario 
is that the bubbles are produced with little kinetic energy. 
For such bubbles there is a relation between their mass and radius, given 
by eq. (\ref{square}) with $\dot{R}=0$. 
For $\ex_1=1$, the resulting function $M(R,m)$ has a maximum at $R_{\rm max}$, corresponding to the critical bubble. Bubbles produced with $R>R_{\rm max}$
and $\dot{R}=0$ can only grow, while bubbles produced with $R<R_{\rm max}$
and $\dot{R}=0$ can only shrink. The value of $M(R_{\rm max},m)$ can be used in
order to estimate
the energy barrier that must be overcome for growing bubbles
to exist. It can be checked that 
the configurations with $\ex_1=-1$ in eq. (\ref{square}) correspond to
shrinking bubbles. (In fig. \ref{fig1} we depict this branch as well, even though
it is not relevant for our discussion below.) 

For a central black hole with mass parameter $m$, 
the presence of
an initially static AdS bubble of radius $R$ 
results in the modification of the asymptotic ADM mass by an amount 
equal to 
$\delta M(R,m)=M(R,m)-m$. In fig. \ref{fig1} we depict this function 
in units of $\ell_{\rm AdS}$ (i.e. $\ell_{\rm AdS}=1$) for a
Schwarzschild exterior ($1/\ell_{\rm dS}\to 0$) and $\kappa=0.1$. 
For $m=0$ the maximum corresponding to the critical bubble is clearly visible.
It is also apparent that very large bubbles have $M<0$ because of the
negative volume contribution. The presence of a central black hole with $m>0$
leads to the increase of the mass $M$ of the total configuration. 
However, the mass difference $\delta M(R,m)$ is a decreasing function of $m$.
There is a critical value
$m_{\rm cr}$, above which $\delta M(R,m)$ is negative for all $R$. This value 
and the corresponding bubble radius 
$R_{\rm cr}$ can be obtained by requiring that  
$\delta M(R_{\rm cr},m_{\rm cr})=\partial \delta M(R_{\rm cr},m_{\rm cr})/\partial R=0$.
We obtain
\be
Gm_{\rm cr}=\frac{1}{3}R_{\rm cr}=\frac{2}{3\sqrt{3}}
\frac{\kx}{\left[\frac{1}{\ell_{\rm dS}^2}+(\frac{1}{\ell_{\rm AdS}}+\kx)^2 \right]^{\frac{1}{2} } \left[\frac{1}{\ell_{\rm dS}^2}+(\frac{1}{\ell_{\rm AdS}}-\kx)^2 \right]^{\frac{1}{2} }}
\simeq \frac{2}{3\sqrt{3}}
\frac{\kx}{\frac{1}{\ell_{\rm AdS}^2}+\frac{1}{\ell_{\rm dS}^2} } .
\label{mcr} \ee
The bubble radius $R_{\rm cr}$ is always larger than the horizon radius
of the black hole.

Another characteristic feature of fig. \ref{fig1} is the 
absence of bubbles with radii below a certain value. 
The perusal of eq. (\ref{square}) reveals that, for $\dot{R}=0$, the minimal radius $R_{\rm h}$ satisfies
\be
1-\frac{2G m}{R_{\rm h}}+\frac{R_{\rm h}^2}{\ell_{\rm AdS}^2}=0.
\label{crr} \ee
It is clear that $R_{\rm h}$ coincides with the horizon of the
AdS black hole of mass $m$. If the bubble is located within the horizon,
with vanishing wall velocity, it cannot grow, as 
the attraction of the wall by the 
black hole is too strong.
For $2 Gm/\ell_{\rm AdS}\ll 1$ we have $R_{\rm h} \simeq 2 G m$,
while for $2 Gm/\ell_{\rm AdS}\gg 1$ we have $R_{\rm h} \simeq (2 G m \ell_{\rm AdS}^2)^{1/3}$.
A significant outward wall velocity allows
bubbles with initial $R<R_{\rm h}$ to grow. However, we expect that the 
majority of bubbles are produced with small velocity.

\begin{figure}
\includegraphics[width=0.7\textwidth]{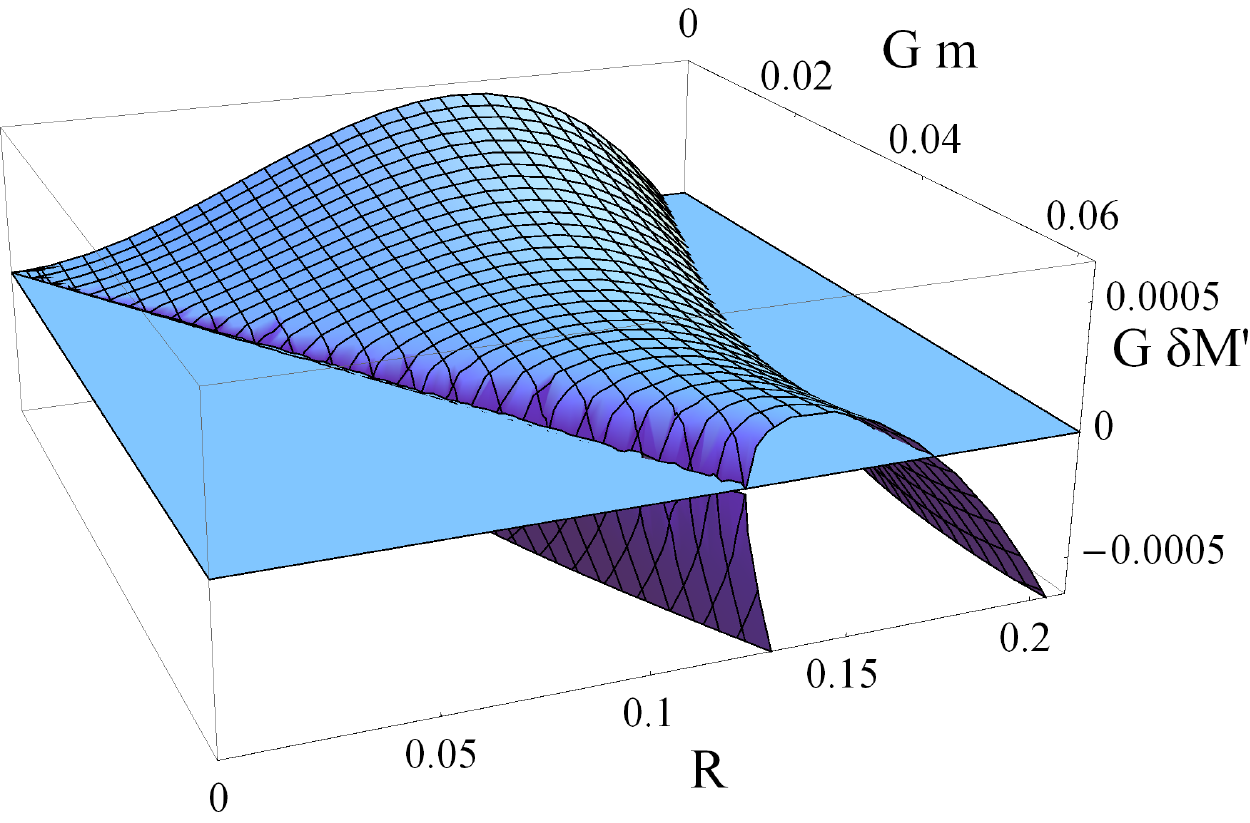}
\caption{The mass difference $\delta M'=M-R_{\rm h}/(2G)$ for a bubble with $\dot{R}=0$ and $\kx=0.1$, $1/\ell_{\rm dS}\to 0$, 
as function of its radius $R$ and the mass $m$ of the central black hole. All quantities are given in units of $\ell_{\rm AdS}$.}
\label{fig2}
\end{figure}

The above analysis indicates that 
a black hole with mass larger than $m_{\rm cr}$ in 
asymptotically flat or dS space may not be 
separated by an energy barrier from a configuration in which it is
surrounded by an AdS bubble. However, the analysis cannot be considered
rigorous because the mass parameter $m$ is not a properly defined physical
quantity. A geometrical quantity that can be used to 
characterize the energy content of the central region is the 
horizon radius $R_{\rm h}$. The difference $\delta M'(R,m)=M(R,m)-R_{\rm h}(m)/(2G)$
provides an alternative estimate for the energy barrier associated with
the bubble. Notice that $R_{\rm h}(m)/(2G)$ coincides with $m$ only for 
$2 Gm/\ell_{\rm AdS}\ll 1$, while it is much smaller than $m$ for
$2 Gm/\ell_{\rm AdS}\gg 1$. As a result, $\delta M'$ may provide an overestimate
of the energy barrier for large $m$.

In fig. \ref{fig2} we depict the function  $\delta M'(R,m)$
for bubbles with $\dot{R}=0$ and $\ell_{\rm AdS}=1$, $\kx=0.1$, $1/\ell_{\rm dS}\to 0$.
We observe the presence of a critical configuration for each value of $m$, 
similarly to fig. \ref{fig1}. The critical of value of $\delta M'$ 
initially drops as a function of $m$, but starts growing again for 
large $m$. The minimal critical value is obtained for 
values ($R'_{\rm cr}, m'_{\rm cr}$) such that 
$\partial \delta M'(R'_{\rm cr},m'_{\rm cr})/\partial R=
\partial \delta M'(R'_{\rm cr},m'_{\rm cr})/\partial m=0$.
An analytical expression for the solution of these equations can be 
obtained for small $\kx \ell_{\rm AdS}$. The full expression for nonzero $1/\ell_{\rm dS}$ 
is lengthy and not very illuminating. For this reason we present only
the solution for $1/\ell_{\rm dS}\to 0$, which reads
\be
Gm'_{\rm cr}=
\frac{ 1+\sqrt{13}}{12} R'_{\rm cr}=
\frac{\sqrt{16-\sqrt{13}}\left( 1+\sqrt{13} \right) }{36} \ell_{\rm AdS}^2 \kx.
\label{mcrpr} \ee
The points ($R_{\rm cr}, m_{\rm cr}$), ($R'_{\rm cr}, m'_{\rm cr}$)
are close, as can be seen from their approximate values:
$Gm_{\rm cr}=R_{\rm cr}/3\simeq 0.385\, \ell_{\rm AdS}^2 \kx$ and 
$Gm'_{\rm cr}=1.151\, R_{\rm cr}/3\simeq 0.450\, \ell_{\rm AdS}^2 \kx$.

The quantity $\delta M'(R'_{\rm cr}, m'_{\rm cr})$ provides an
estimate of the minimal energy barrier associated with the AdS bubble. 
For $1/\ell_{\rm dS}\ll 1/\ell_{\rm AdS}$
we find a positive value, which can 
be compared with the barrier in the absence of the black hole,
estimated by $M(R_0,0)$, with $\partial M(R_0,0)/\partial R=0$. 
For small $\kx \ell_{\rm AdS}$ and $1/\ell_{\rm dS} \to 0$, we find
\be
\frac{\delta M'(R'_{\rm cr}, m'_{\rm cr})}{M(R_0,0)}=
\frac{1}{32}\left( 
\sqrt{220+47\sqrt{13}}-\sqrt{1492-397\sqrt{13}}  \right)
\simeq 0.373. 
\label{barrier} \ee
We have verified this value through a numerical solution.
The more conservative estimate of the energy barrier by 
$\delta M'$ does not indicate complete instability. 
However, the presence of a black hole has a 
profound influence, as it can reduce
the energy barrier by a factor of roughly 3. This has a strong effect on
the nucleation rate, which is exponentially suppressed by the
the energy barrier.

For $1/\ell_{\rm dS}\gg 1/\ell_{\rm AdS}$ we have seen 
that $\delta M$ for the critical 
bubble turns negative for a central black hole with a mass parameter
given by eq. (\ref{mcr}): $Gm_{\rm cr} \sim \kappa \ell_{\rm dS}^2$.
As in this case $2Gm_{\rm cr} \ll \ell_{\rm AdS}$, we have $2Gm_{\rm cr}\simeq R_{\rm h}$. 
This means that $\delta M'(R,m)$ also turns negative for a central 
black hole with this mass parameter, so that complete instability is expected
for a strong dS background. 

The fact that $m_{\rm cr}$, $m'_{\rm cr}$ are roughly equal for $1/\ell_{\rm dS}\to 0$,
while for $1/\ell_{\rm dS}\gg 1/\ell_{\rm AdS}$ the instability appears for a black-hole mass 
given by eq. (\ref{mcr}), leads to the conclusion that the critical mass can
be estimated through the relation
\be
Gm_{\rm cr} \sim \frac{\kx}{\frac{1}{\ell_{\rm AdS}^2}+\frac{1}{\ell_{\rm dS}^2} }
\label{criticalmass} \ee
for all values of $1/\ell_{\rm AdS}$, $1/\ell_{\rm dS}$.


\section{Higgs (in)stability}\label{estimate}

In order to determine the relevance of our analysis for the 
fate of the Higgs field, we must recall 
some results
for the Higgs potential in the Standard Model \cite{tetradis}:
The Higgs potential has the approximate form $V \sim \lx(h)\, h^4/4$ for 
values of the Higgs field $h$
above $10^6$ GeV. The quartic coupling $\lx$ varies from 0.02 to $-0.02$ 
for Higgs values between $10^{6}$ GeV and $10^{20}$ GeV, respectively. 
The maximum of the potential is located at a value $h_{\rm max}\sim 5\times
 10^{10}$ GeV. 
 Near the maximum the potential can be approximated as \cite{tetradis}
 \be
 V(h)\simeq -b\ln \left( \frac{h^2}{h^2_{\rm max}\sqrt{e}} \right)\frac{h^4}{4},
 \label{apprvm} \ee
 with $b\simeq 0.16/(4 \pi)^2$. For Higgs values within the range of interest around
 $h_{\rm max}$, we have $|\lx|={\cal O}(10^{-3})$.
 In order to avoid the destabilization of the standard electroweak vacuum
 because of Higgs fluctuations during inflation, 
 one requires that the scale $H_{\rm inf}$ of inflation
 satisfies $H_{\rm inf}\lta 0.04\, h_{\rm max}$ (for a minimally coupled Higgs field). 
 When this constraint is satisfied the probability for the Higgs field to 
 fluctuate beyond the maximum of the potnetial is exponentially small.
 
 The presence of a black hole alters this picture drastically. If its mass
 is comparable to the critical value (\ref{criticalmass}), the energy barrier
 to be overcome in order to produce an AdS bubble around a black hole
 is reduced significantly
 compared to the barrier in the absence of a black hole.
Masses larger than $m_{\rm cr}$ are also relevant, because Hawking 
evaporation makes the black-hole mass scan a wide range of values during the cosmological evolution. 
  It is, therefore,
 important to compare the specific value of $m_{\rm cr}$ for the Higgs
 potential with the maximal mass $m_{\rm bh}\sim \mpl^2/H$ of primordial
 black holes
 produced at some given time during or after inflation. 
A precise determination of the bubble profile requires a
numerical solution of the equation of motion for the Higgs field. 
Before presenting this solution, we discuss
an estimate based on the thin-wall approximation of the previous
sections, omitting factors of order 1.

 We use rough estimates of the Higgs potential near $h_{\rm max}$ in terms
 of a positive parameter $\lh={\cal O}(10^{-3})$.
 The energy density in the interior of a bubble is $V\sim - \lh h^4$.
 The surface tension can be estimated from the Newtonian expression 
 \be \sigma 
 \sim \frac{h^2}{\Delta r} + \Delta r \lh h^4 \gta 
 \sqrt{\lh} h^3,
 \label{surface}\ee
 minimized for a bubble thickness $\Delta r \sim 1/
 \left( h\sqrt{\lh}\right)$.
The critical value (\ref{criticalmass}) can now be compared to the maximal
mass $m_{\rm bh}\sim \mpl^2/H$ of a primordial black hole produced at 
a given time. We obtain
\be
\frac{m_{\rm cr}}{m_{\rm bh}}\sim  \frac{|V|}{V'+|V|}\, \frac{H}{\sqrt{\lh}h},
\label{res} \ee
with $V\sim -\lh h^4$, and $V'$ equal to the inflaton vacuum energy $V_{\rm dS}$ during inflation, or to zero after its end. 
\begin{itemize}
\item
During inflation, the constraint 
$H_{\rm inf}\lta 0.04\, h_{\rm max}$ gives 
$H_{\rm inf}/(\sqrt{\lh} h_{\rm max}) \lta 1$
for  $\lh={\cal O}(10^{-3})$, while we must have
$|V|/V' \ll 1$ for the positive vacuum energy to dominate.
Thus we obtain
\be
\frac{m_{\rm cr}}{m_{\rm bh}}\lta  \frac{|V|}{V'}\, \frac{h_{\rm max}}{h}.
\label{res1} \ee
For Higgs values within the bubble 
in the vicinity of $h_{\rm max}$, 
the mass of the typical black hole is larger than the critical mass for
the instability to appear. 
\item
After the end of inflation, $V'=0$ and $H^2\sim \lh h^4/\mpl^2$, so that
\be
\frac{m_{\rm cr}}{m_{\rm bh}} \sim \frac{h}{\mpl}
\label{res11} \ee
\end{itemize}
and the maximal black hole mass is larger than $m_{\rm cr}$.
Black holes with masses that are a fraction of the maximal value
fall within the region that induces the 
drastic reduction of the energy barrier. 

Despite the absence of a significant energy barrier, the formation of an AdS 
bubble cannot proceed if its typical size is larger than the size of the
causally connected regions in the Universe $R_{\rm H}\sim 1/H$. 
From eq. (\ref{mcr}) we have
\be
\frac{R_{\rm cr}}{R_{\rm H}}\sim  \frac{|V|}{V'+|V|} \frac{H}{\sqrt{\lh}h}.
\label{ress} \ee
As we discussed above, this ratio is smaller than one at all times. 

\section{Numerical solutions}\label{numerical}

The analysis of the previous sections is based on the thin-wall 
approximation, in which the transition from the interior AdS-Schwarzschild
spacetime to the external asymptotically flat or dS spacetime takes place
within a shell of very small width. Moreover, the vacuum energy is
assumed to be constant on either side of the wall. In the realistic
scenario, the Higgs field varies continuously as a function of the radius
and the bubbles have large thickness. 
The profile of the critical bubble can be obtained by solving the equation
of motion of the Higgs field with appropriate boundary conditions 
that force the field to interpolate between values on either side of
the maximum of the potential. As the critical bubble is static (but unstable)
the corresponding field configuration has no time dependence. 
The ADM mass for this configuration quantifies the energy barrier that
must be overcome for the field to fluctuate beyond the potential maximum.
The nucleation rate is expected to be suppressed by this mass. 
The characteristic scale of the fluctuations, which determines the
dispersion of the Higgs field, is set by the environment. If
thermal equilibrium is assumed, the scale is given by the temperature.
However, a high density environment with strong fluctuations also
affects the Higgs field through the Yukawa couplings to the particles
that contribute to the density.

\begin{figure}[!t]
\centering
$$
\includegraphics[width=0.45\textwidth]{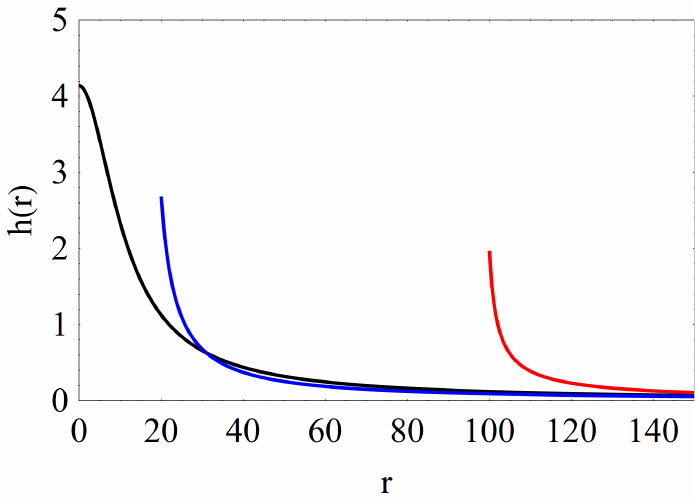}\qquad 
\includegraphics[width=0.45\textwidth]{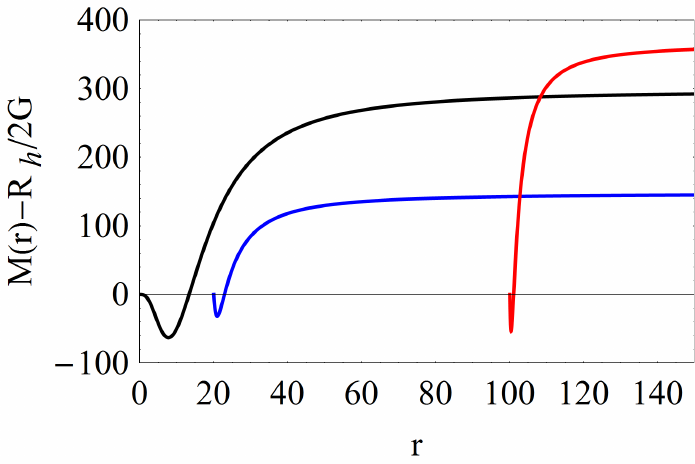}\qquad $$
\caption{\em The Higgs field $h(r)$ (left plot) and the mass function
$M(r)$ (right plot) outside a black hole with horizon
radius $R_{\rm h}=$0.1, 20 and 100 
(lines from from left to right). All quantities are given 
in units of $h_{\rm max}$.  
}
\label{higgsmass}
\end{figure}

The purpose of this first study is not to analyze a specific scenario, but 
to establish the effect of a black-hole background on the nucleation rate.
For this reason, we concentrate on the role of black holes
after the end of inflation, assuming an asymptotically flat spacetime. 
We also employ the zero-temperature Higgs potential, neglecting
temperature or density corrections. A more focused analysis will be
presented in the future. 
We solve numerically the Einstein
and Higgs-field equations for the static critical bubbles, 
using the potential of eq. (\ref{apprvm}).
We assume a metric of the form
\be
ds^2=-N(r) \, e^{2\delta(r)} dt^2+N^{-1}(r)\, dr^2
+r^2 \left(d\theta^2+\sin^2\theta d\phi^2 \right),
\label{metric} \ee 
with $N(r)=1-2G M(r)/r$. 
The equations of motion are reduced to 
\begin{eqnarray}
M'&=&4\pi r^2 \left(\frac{1}{2} N h'^2+V(h) \right)
\label{eom1} \\
\delta'&=&4\pi G r h'^2
\label{eom2} \\
h''+\left( \frac{2}{r} -8\pi G\frac{r}{N}V(h)+2G\frac{M}{Nr^2}
\right) h'&=&\frac{1}{N}\frac{dV(h)}{dh},
\label{eom3}
\end{eqnarray}
with the prime denoting a derivative with respect to $r$.

The presence of the horizon at $r=R_{\rm h}$ requires the introduction of
appropriate boundary conditions at this point. The function $N(r)$ vanishes
at $r=R_{\rm h}$, which sets 
\be
2GM(R_{\rm h})=R_{\rm h}.
\label{bound1} \ee 
In order to obtain a finite value for the ADM mass of the configuration 
we must avoid a singularity in eq.
(\ref{eom3}) at $r=R_{\rm h}$. This can be achieved only if we impose 
\be
h'(R_{\rm h})= \frac{R_{\rm h}}{1-8\pi G R^2_{\rm h}\, V(h(R_{\rm h}))}
 \frac{dV(h(R_{\rm h}))}{dh}.
\label{bound2} \ee
Notice that this boundary condition correctly reproduces the 
standard condition $h'(0)=0$ in the absence of a black hole.
We also require that 
\begin{eqnarray}
h(r)&\to& 0~~~~~~~~~~~~~~~{\rm for}~r\to \infty
\label{bound3} \\
\delta(r)&\to& 1~~~~~~~~~~~~~~~{\rm for}~r\to \infty.
\end{eqnarray}
With these conditions $M(r)$ corresponds to the mass function of eq. (\ref{square}) in the thin-wall approximation. We are interested in
the quantity $\delta M'=M(\infty)-R_{\rm h}/(2G)$, which corresponds to the
similarly denoted quantity of section \ref{central}. It can be seen
from eqs. (\ref{eom1}), (\ref{bound1}) that $\delta M'$ results from the 
integration of the energy contributions associated with the bubble. It
gives an estimate of the barrier 
associated with the production of the bubble around
a central black hole with horizon radius $R_{\rm h}$. 

For the 
potential of eq. (\ref{apprvm}),
the set of eqs. (\ref{eom1})-(\ref{eom2}) has the trivial solution $h(r)=0$, 
$M(r)=R_{\rm h}/(2G)$, and a nontrivial one with
a characteristic scale 
set by $h_{\rm max}\sim 5\times 10^{10}$ GeV. The hierarchy
between this scale and $\mpl$ makes the rhs of eq. (\ref{eom2}) negligible.
As a result, we have $\delta=1$ to a very good approximation in all the
solutions we present. 

In fig. \ref{higgsmass} we depict the solutions for 
the Higgs field $h(r)$ (left plot) and the quantity
$M(r)-R_{\rm h}/(2G)$ (right plot) in the exterior of the black hole.
We present three solutions for black holes with horizon
radii $R_{\rm h}=$0.1, 20 and 100 
(lines from from left to right). All quantities are given 
in units of $h_{\rm max}$.  
For bigger black holes the derivative of the
Higgs field at $r=R_{\rm h}$ tends to become more negative (see eq. (\ref{bound2})).
As a result, the Higgs can reach the origin starting 
from smaller initial values: the transition to the vicinity of the origin
is faster for larger $R_{\rm h}$. The quantity $M(r)-R_{\rm h}/(2G)$ is zero on the
horizon, becomes negative initially because of the negative contribution 
from the potential, but quickly turns positive as the field moves through 
the maximum of the potential. For $r\to \infty$ it approaches a constant value $\delta M'$. 
This asymptotic value has a dependence on $R_{\rm h}$ similar to the one we
discussed in section \ref{central}. For $R_{\rm h} \to 0$ we have 
$\delta M'=M_0$, with $M_0$ denoting the size of the energy barrier in the absence of a black hole. For nonzero $R_{\rm h}$, 
$\delta M'$ initially decreases, as can be seen in fig. \ref{higgsmass} for
$R_{\rm h}=20$, but eventually grows, as can be seen for $R_{\rm h}=100$.

\begin{figure}
\includegraphics[width=0.7\textwidth]{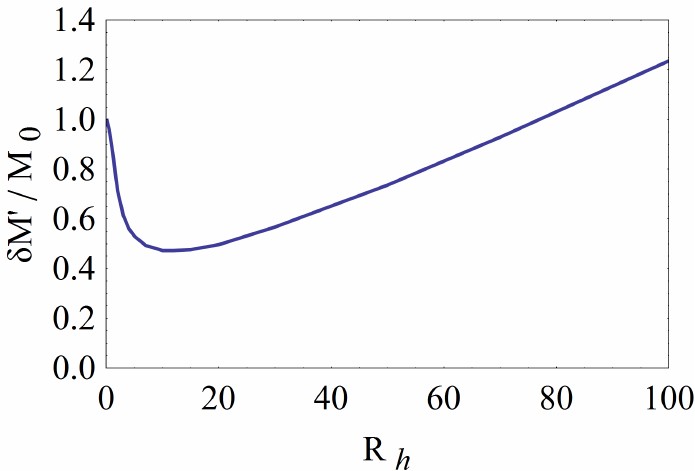}
\caption{The ratio of the energy barrier
in the background of a black hole relative to the barrier in the absence 
of a black hole. }
\label{ratio}
\end{figure}

The ratio $\delta M'/M_0$ denotes the relative size of the energy barrier
in the background of a black hole with respect to the barrier in its absence. 
We depict it in fig. \ref{ratio}. We observe a minimum
$\delta M'_{\rm min}\simeq 0.473$ at $R_{\rm h}\simeq 11$.
This value can be compared to the result (\ref{barrier}), derived through
the thin-wall approximation. The small discrepancy is expected, as the 
numerical solutions for $R_{\rm h}\sim 10$
correspond to bubbles with significant wall thickness. There is a reduction
of the energy barrier by approximately a factor of 2, instead of 
the factor of 3 indicated by eq. (\ref{barrier}). However, this reduction still
has a profound effect on the nucleation rate. 

We conclude this section with some remarks on the properties 
of the solutions we presented:
\begin{itemize}
\item
The characteristic scale of the solutions is set by $h_{\rm max}$. This means
that gravitational corrections are not relevant, as they are suppressed
by powers of $h^2_{\rm max}/\mpl^2$. Modifications are possible only if the
Higgs potential receives corrections from new physics at 
scales close to $h_{\rm max}\sim 10^{10-11}$ GeV.
\item
For our solutions 
$\kx \ell_{\rm AdS} \sim \sqrt{G} h_{\rm max} \sim h_{\rm max}/\mpl \ll 1$. 
We have used this approximation in order to derive the various estimates 
in section \ref{central}. 
\item
The energy $\delta M'$ associated with 
the bubble is much smaller than the mass of the central black hole, as 
estimated by $R_{\rm h}/(2G)$. 
This can be seen from the ratio $G\, \delta M'/R_{\rm h}\sim h^2_{\rm max}/\mpl^2$.
As a result, the gravitational background is induced mainly by the black hole, 
with the bubble being only a small perturbation. This is the reason why the 
metric parameter $\delta(r)$ equals 1 to a very good approximation.
\end{itemize}

\section{Summary and conclusions}\label{conclusions}

The general conclusion that can be reached through the analysis of sections
\ref{AdS} and \ref{central} is that
the presence of a black hole reduces the energy barrier for the
destabilization of a false vacuum through classical fluctuations. The 
magnitude of the effect for the transition to an AdS vacuum depends on 
the quantity that one associates with the barrier. One may employ the
difference $\delta M = M-m$ 
between the ADM mass $M$ of the black hole-bubble configuration
and the mass parameter $m$ of the black hole. In this case the barrier 
is completely eliminated for sufficiently large $M$, indicating complete
instability. However, a quantity with a clearer physical meaning is 
the difference  $\delta M' = M-R_{\rm h}/(2G)$, with $R_{\rm h}$ the horizon radius.
If $\delta M'$ is employed in order to estimate the energy barrier, one
must distinguish two cases: a) For an {\it asymptotically flat false vacuum},
the barrier is not completely eliminated in the presence of
a black hole However, an analysis based on
the thin-wall approximation indicates that the barrier
can be reduced by a factor of roughly 3.
We point out that $R_{\rm h}/(2G)$ coincides with $m$ only for 
$2 Gm/\ell_{\rm AdS}\ll 1$, with $\ell_{\rm AdS}$ the AdS length,
while it is much smaller than $m$ for
$2 Gm/\ell_{\rm AdS}\gg 1$. As a result, $\delta M'$ may provide an overestimate
of the energy barrier for large $m$. b) For an {\it asympotically dS false 
vacuum}, $\delta M'$ turns negative for a sufficiently big black hole,
indicating a complete instability. 

The discussion of sections \ref{AdS} and \ref{central} is not specific to
the potential of the Higgs field, as it is formulated 
only in terms of the energy density inside and outside 
the bubble, and the
surface tension of the wall. When these are estimated from the 
Higgs potential (section \ref{estimate}), it becomes apparent that the 
presence of a typical primordial black hole has a profound effect 
on the energy barrier for classical transitions towards the 
region in which the potential becomes negative. 
The same is expected to be true for a wide range of potentials, 
resulting from physics beyond the Standard Model at zero and nonzero 
temperature, as long as the standard electroweak vacuum is metastable.

The calculation in section \ref{numerical} of the exact bubble profile and the corresponding energy barrier for the 
Standard-Model Higgs in asymptotically flat space confirms the
general picture, but also provides an estimate of the level of 
accuracy of the analysis of the earlier sections.
In particular, it indicates that 
$\delta M'$ can be reduced by an approximate factor of 2, instead of
the approximate factor of 3 found in section \ref{central}. Even with this
correction, the energy of the barrier $\delta M'$, when computed through the
zero-temperature potential, drops from $\sim 300$ to $\sim 150$
in units of the Higgs value $h_{\rm max}$ at the maximum of the potential.
The bubble nucleation rate per unit time
scales proportionally to $\exp(-\delta M'/T)$. For temperatures
$T\sim h_{\rm max}$, for which the temperature corrections do not
dominate the zero-temperature potential, 
the effect can be dramatic, with this factor 
being reduced from $10^{-130}$ to $10^{-65}$. An analogous effect 
is expected at higher temperatures as well, but a precise analysis
requires the use of the temperature-corrected potential.

The scenario in which 
primordial black holes, produced after inflation, 
induce the nucleation of bubbles of
AdS vacuum is phenomenologically untenable.
The expansion of the bubbles leads to catastrophic crunch 
singularities (see ref. \cite{tetradis} and appendices \ref{app2} and \ref{app3}), 
so that our visible Universe cannot develop.
Whether this catastrophic scenario can be avoided 
depends strongly on the precise value of the energy barrier, but also on 
several other model-dependent factors, such as 
the rate of production of black holes and their typical mass.

 The usual assumption is that primordial black holes can form after inflation 
 when certain horizon-size overdense regions collapse gravitationally. The number
of causally independent regions, that may have collapsed 
and are currently within our horizon, can be huge. The volume of our present 
horizon $(3.4/H_0)^3$ gets suppressed by $a^3$ at earlier times when
the scale factor is $a <1$. On the other hand, the volume of a horizon-size
region during the radiation era is $1/H^3$. Therefore, the number of
causally independent regions is roughly $N\simeq a^3 (3.4 H/H_0)^3$.
The ratio $H/H_0$ can be expressed 
as $H/H_0=\sqrt{g_* \Omega_\gamma/2}(T/T_0)^2$, where $g_*=106.75$ 
is the number of degrees of freedom of the Standard Model, 
$T_0\simeq 2.4 \times 10^{-13}$ GeV the current photon temperature, and 
$\Omega_\gamma \simeq 5 \times 10^{-5}$ the current energy fraction in photons. 
The scale factor $a$ can be related to the temperature through 
entropy conservation: $g_*T^3 a^3=g_{*S0}T^3_0$, with $g_{*S0}=3.94$ the 
number of degrees of freedom contributing to the entropy after 
$e^+e^-$ ahhihilation. Putting everything together gives
$N \sim 10^{34} (T/{\rm GeV})^3. $
At a temperature $T\sim h_{\rm max}\sim 10^{11}$ GeV, we have $N\sim 10^{67}\sim \exp(154)$. If the primordial 
black holes are produced with sufficient probability, there is room for
their effect to compensate the exponential suppression of the bubble
nucleation rate.  
It should also be noted that Hawking evaporation 
makes the black-hole mass scan a wide range of values during the cosmological
evolution, so that the energy barrier for the nucleation of AdS bubbles
may vary with time through its minimal value.  

The nucleation rate per unit time in the background of 
one black hole is $dP/dt \sim T \exp(-\delta M'/T)$.
Notice that there is no volume factor, as the presence of the black hole
eliminates the translational invariance. 
The characteristic time interval that can be associated with the 
scale $T$ is the Hubble time $1/H\sim \mpl/T^2$. 
The nucleation is efficient over longer times, but we are interested
in a lower bound for the probability. 
Neglecting the evaporation of the
black hole, we have a bubble-nucleation probability 
$P\sim \mpl/T \exp(-\delta M'/T)$. 
For $N$ causally independent regions, in each of which a primordial black 
hole can appear with probability $p$, the total nucleation probability
becomes $P\sim N p \mpl/T \exp(-\delta M'/T)\sim p\, \left( T/10^{11}\right)^2\,\exp(173-\delta M'/T)$. 
When the barrier is reduced  to $\delta M'/T \simeq 150$, as we discussed
earlier, the 
exponential suppression is eliminated. It must be emphasized, however,
that the probability $p$ for the creation of a primordial black hole may be
very small, resulting in the suppression of the rate.

It is clear from the above that 
the analysis of a detailed scenario requires the precise
determination of several physical parameters. In this work we identified 
a crucial factor in this discussion, namely
the effect of the black-hole background on the height of the energy
barrier for transitions to an AdS vacuum. 
In the most typical case, the transition to the AdS vacuum is induced by
fluctuations within a Universe in thermal equilibrium. 
One may ask if the black holes that affect vacuum stability 
can be in equilibrium with the thermal
background or affect it. For the black-hole masses of interest, this does not
seem to be the case. As can be seen from fig. \ref{ratio}, the tunnelling
rate becomes maximal for a black hole with Schwarzschild radius 
$R_{\rm h}\simeq 10/ h_{\rm max}$ and Hawking temperature 
$T_{\rm H}\simeq h_{\rm max}/(40 \pi)$. We expect that the bubble nucleation 
rate will be most efficient for $T\gta h_{\rm max}$. This means that the
black holes are not in equilibrium, while they
have only minor influence on the background. As a result, the
main effect of the black holes on the critical bubbles is gravitational.

It must be kept in mind also that the Higgs field
couples through the Yukawa couplings to particles that 
contribute to density flucuations. This implies that 
the transition can take place in an environment that is out of thermal equilibrium if the 
density flucuations are sufficiently strong. 
The temperature, or the Higgs-field dispersion induced by density perturbations,
are additional model-dependent parameters whose time evolution 
must be calculated precisely
in order to determine if the production of primordial 
black holes is consistent with our observable Universe.

\subsubsection*{Acknowledgments}
I would like to thank G. Giudice, A. Salvio, A. Strumia, A. Urbano 
for useful discussions.



\appendix

\section{Equation of motion}\label{app1}

By squaring eq.~(\ref{junction12}) twice,
we can express the result as the equation
for the one-dimensional motion of a particle of total `energy' $E$ in a `potential' $V$:
\be
\left(\frac{d\Rt}{d\ttau} \right)^2+V(\Rt)=E,
\label{eom} \ee
with
\be
V(\Rt)=-\left( \frac{\ex_M-\zeta+\ex \Rt^3}{\Rt^2}\right)^2-\ex_M\frac{\gamma^2}{\Rt}-\delta^2\Rt^2,
\qquad
E=-\frac{\kappa^2}{G^2M^2\rho^4},
\label{energ}\label{pot} \ee
where
\beq 
\Delta=  \frac{1}{\ell_{\rm AdS}^2} +\frac{1}{\ell_{\rm dS}^2}-\kappa^2,
\quad 
\rho^3 = \frac{1}{2G|M|}\left| \Delta \right|,
\quad
\gamma=\frac{2\kappa}{\left| \Delta \right|^{1/2}},
\quad
\kappa^2=\left(\frac{1}{\ell_{\rm AdS}^2}+\frac{1}{\ell_{\rm dS}^2}\right)
\frac{\gamma^2}{\gamma^2+4\ex},
\quad
\delta^2=\frac{4\kx^2}{\ell_{\rm dS}^2\Delta^2},
\label{epss}  
\eeq
and
$\ex = {\rm sign} \left(\Delta \right)$, $\epsilon_M={\rm sign}(M)$,
$\zeta=m/|M|$.
The dimensionless  `coordinate' variable $\Rt$ and the `time' variable $\tilde\tau$ are defined as
$\Rt= \rho R$ and $\ttau=2\kappa\tau/\gamma^2$.

The form of the solutions of eq.\ (\ref{eom}) can be revealed more easily through the following observations:
\begin{itemize}
\item
The sign $\ex_2$ disappeared when performing the second squaring, so that
eq.~(\ref{eom}) describes the solutions of eq.~(\ref{junction12}) with both
values of $\ex_2$. 
We can rewrite eq.~(\ref{junction12}) in terms of the new parameters as
\be
\beta_{\rm in}=\beta_{\rm out}+4\pi G \sigma R = \frac{G|M|\rho^2}{\kappa}\frac{1}{\Rt^2}\left(\ex_M-\zeta+\ex \Rt^3+\frac{\gamma^2}{2}\Rt^3 \right),
\label{cond1} \ee
where we have used eq.~(\ref{square1}). For growing bubbles with 
increasing $\Rt$, we have $\beta_{\rm in}>0$ (i.e. $\ex_1=1$) after a sufficiently long
time. This is obvious for 
$\ex=1$. It also holds for $\ex=-1$ , because 
$\gamma^2>4 $ in this case.
\item
We can also write 
\be
\beta_{\rm out}=\frac{G|M|\rho^2}{\kappa}\frac{1}{\Rt^2} \left(\ex_M-\zeta+\ex \Rt^3 \right),
\label{cond2} \ee
from which it is apparent that, for growing bubbles at late times, we have
$\beta_{\rm out}>0$ (i.e. $\ex_2=1$)
for $\ex=1$, while $\beta_{\rm out}<0$ (i.e. $\ex_2=-1$) for $\ex=-1$. 


\item
For fixed $\ell_{\rm AdS}$, $\ell_{\rm dS}$ and $\kappa$, the total energy $E$ is a function of $M$. As a result, the nature of
the various
solutions of eq.~(\ref{eom}) is directly related to the mass of the bubble. 

\item
The `potential' has a maximum at $\Rt=\Rt_{\rm max}$, given by
\be
\Rt_{\rm max}^3= 
\frac{\ex_M(2\ex+\gamma^2)-2\ex \zeta+\sqrt{(\ex_M(2\ex+ \gamma^2)-2\ex\zeta)^2+32(1+\delta^2)(\ex_M-\zeta)^2}}{4(1+\delta^2)}.
\label{zmm} \ee
In certain cases (e.g. for $\ex_M=-1$, $\delta=\zeta=0$) 
the maximum of the `potential' can be positive. 
The `potential' then vanishes at 
\be
\Rt_{1,2}^3= 
\frac{-\ex_M(2\ex+\gamma^2)+2\ex\zeta\pm 
\sqrt{\gamma^4+4\ex \gamma^2 (1-\ex_M \zeta)-4\delta^2(\ex_M-\zeta)^2}}{2(1+\delta^2)}.
\label{z0ab} \ee

\item
For $\ex_M=1$, the horizon of the black hole of mass $M$ seen by the external observer corresponds to a value $\Rt_{\rm H}$ such that 
\be
E=-\frac{\gamma^2}{\Rt_{\rm H}}-\delta^2\Rt_{\rm H}^2.
\label{bbb} \ee
This relation determines the location of the horizon on a solution of eq.
(\ref{eom}) with given $E$. 
Making use of the definition (\ref{pot}) of the `potential', we can write
\be
E
=V(\Rt_{\rm H})+\left(\frac{1-\zeta+\ex \Rt_{\rm H}^3}{\Rt^2_{\rm H}}   \right)^2.
\label{horiz} \ee
The curve $-\gamma^2/\Rt-\delta^2\Rt^2$, depicting the location of the
horizon, is tangent to the curve $V=V(\Rt)$ at 
\be
\Rt^3=\ex(\zeta-1),
\label{horiii} \ee 
as long
as this is positive. It is apparent from eq. (\ref{cond2}) that $\beta_{\rm out}$ and $\ex_2$ change sign at this point.

\item
The horizon of the AdS black hole of mass $m$ corresponds to a value
$\Rt_{\rm h}$ such that 
\be
E=\left[\frac{\gamma^2(\gamma^2+4\ex)}{4}-\delta^2\right]\Rt_{\rm h}^2-\frac{\zeta \gamma^2}{\Rt_{\rm h}}.
\label{hori} \ee
The curve depicting this horizon is tangent to $V(\Rt)$ at 
\be
\Rt^3=\frac{2(\zeta-\ex_M)}{\gamma^2+2\ex},
\label{sechor} \ee
as long at this is positive. It is apparent from eq. (\ref{cond1}) that
$\beta_{\rm in}$ and $\ex_1$ change sign at this point.

\end{itemize}

\begin{figure}[!t]
\centering
$$
\includegraphics[width=0.45\textwidth]{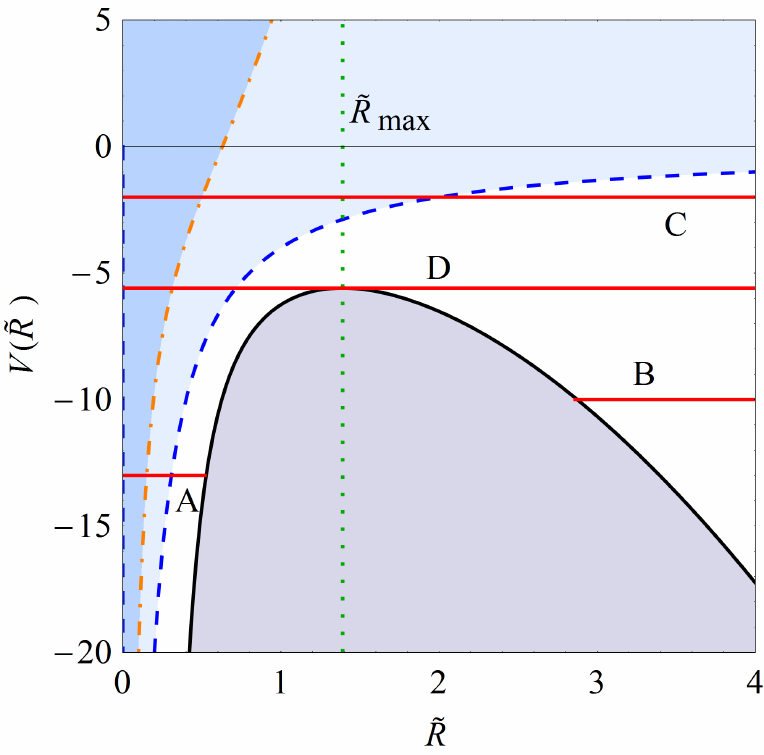}\qquad 
\includegraphics[width=0.45\textwidth]{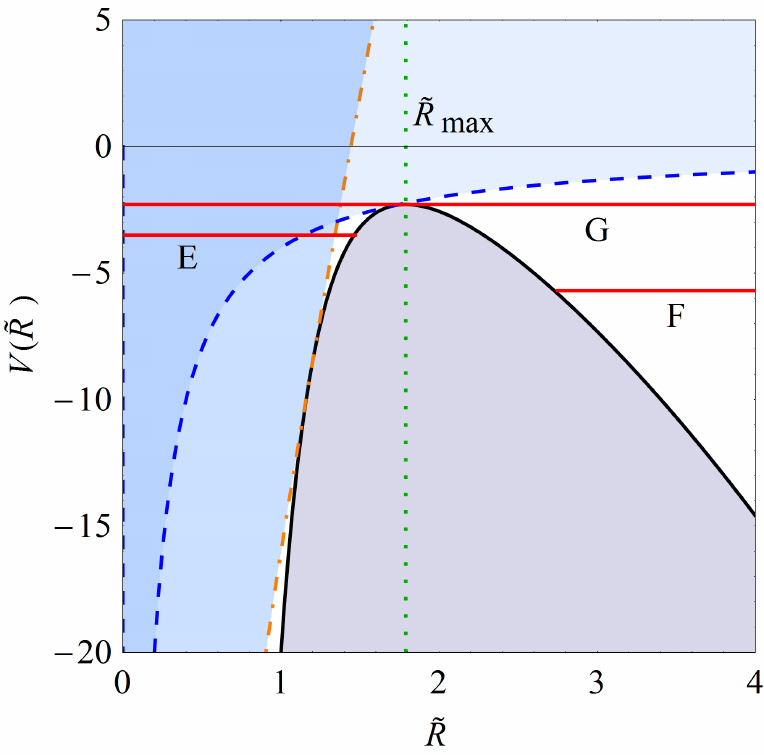}\qquad $$
\caption{\em The `potential' of eq.~(\ref{pot}) for 
$\ex_M=1$, $\gamma=2$, $\ex=1$, $\delta=0$, $\zeta=0.5$ (left), 
and
$\ex_M=1$, $\gamma=2$, $\ex=1$, $\delta=0$, $\zeta=6$ (right).
}
\label{h1}
\end{figure}

\section{Bubble evolution in asymptotically flat space}\label{app2}

We consider first the case of a bubble evolving in asymptically flat
space, which corresponds to $\delta=0$. 
Possible solutions of eq. (\ref{eom}) are depicted in fig. (\ref{h1}) for 
positive bubble mass $M$ $(\ex_M=1)$, and two values of 
the ratio $\zeta=m/M$. We have plotted the `potential' $V(\Rt)$
and indicated the horizons of the internal (dot-dashed line)
and external geometries (dashed line). The straight horizontal lines
depict possible configurations of constant $E$ and, therefore, 
bubble mass $M$ seen by an outside observer. The lines A and E correspond to 
bubbles that expand from vanishing radius to a maximal size and then recollapse.
During their evolution their walls cross the horizons. The lines B and F correspond to
bubbles that start with infinite radius, shrink to a minimal size and then
reexpand. Their walls are located outside both horizons. It is also possible 
for a bubble to be created spontaneously (with the appropriate wall velocity)
at any point along the trajectories, and evolve from there. The lines D and G
correspond to bubbles of critical mass. They have vanishing wall velocity and
a radius corresponding to the location of the maximum of the
`potential'. However, they are unstable, so that small deformations can lead
to their expansion or collapse. The line C correspond to a bubble whose radius
can vary continuously from vanishing to infinite as a function of time.

\begin{figure}[t]
$$\hspace{-1cm}\includegraphics[width=80mm,height=80mm]{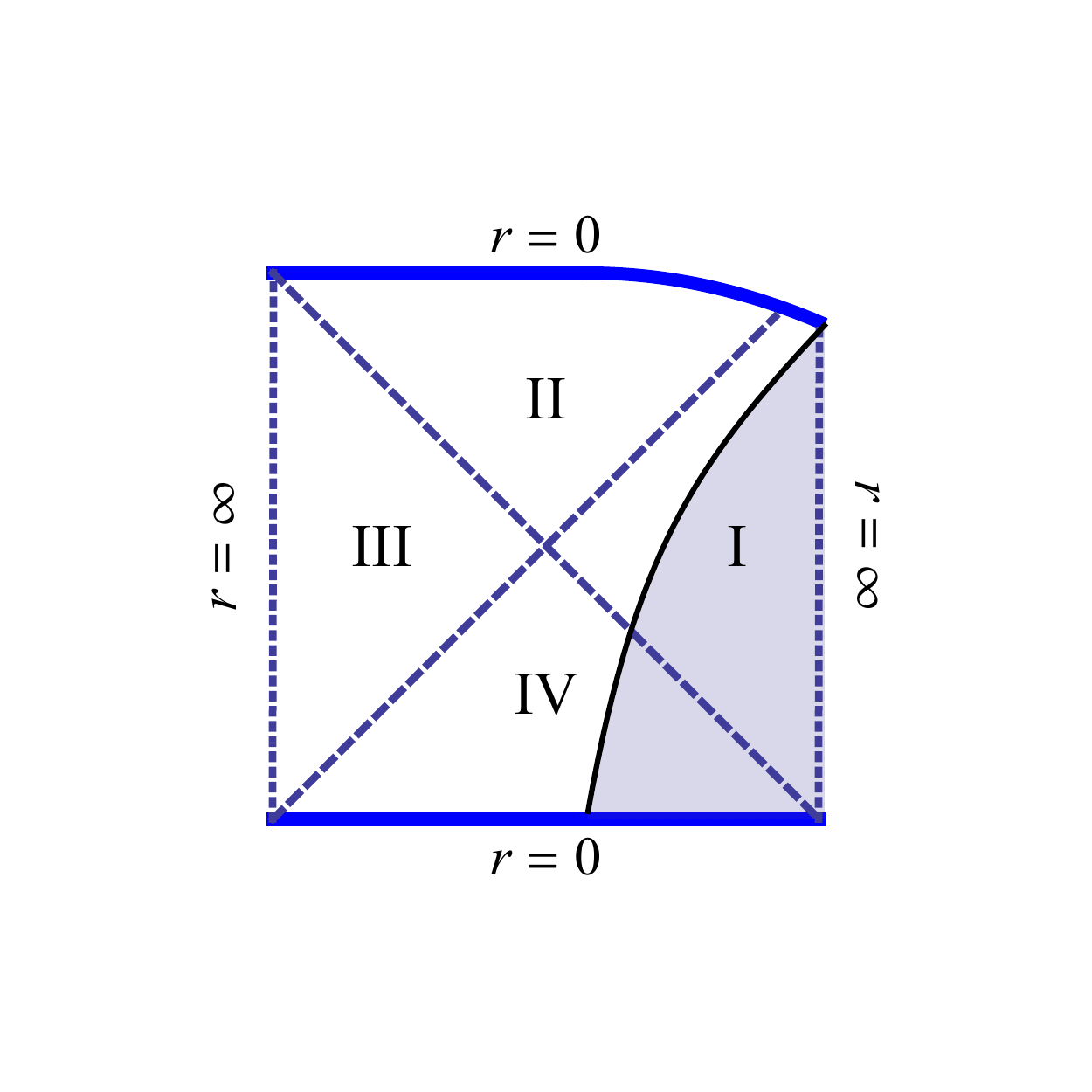}
\includegraphics[width=80mm,height=80mm]{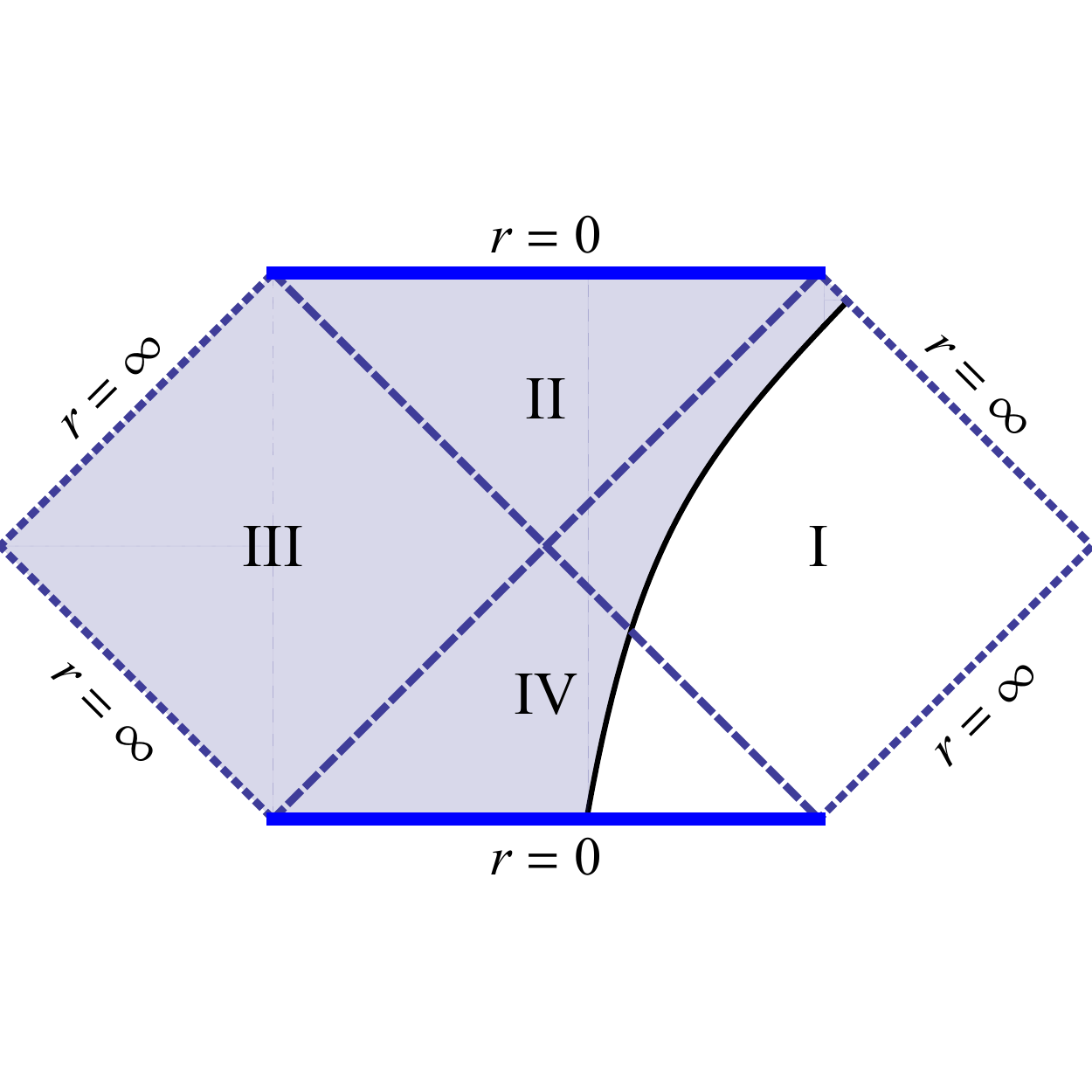} $$
\caption{  \em The Penrose diagram for the wall trajectory C of fig. \ref{h1}. The total spacetime is constructed by joining the two diagrams, after
the elimination of the shaded areas. }
\label{W1}
\end{figure}

Penrose diagrams can be constructed, depicting the full spacetime for 
every type of wall trajectory. A typical example is shown in fig. \ref{W1},
corresponding to the trajectory C of fig. \ref{h1}. 
Thick blue lines denote curvature singularities, 
the dashed lines horizons and the dotted lines conformal infinities.
The total spacetime is 
constructed by eliminating the shaded areas of the two diagrams and joining 
them along the wall trajectory. The bubble wall has initially
vanishing radius, which grows with time. The wall crosses the horizons of 
both the AdS-Schwarzschild (left) and Schwarzschild (right) geometries, and
eventually reaches infinity. 

The AdS interior is expected to be dynamically unstable~\cite{deluccia}.
For $m=0$, the AdS space, when described in FRW coordinates,
displays a coordinate singularity
in a finite amount of internal time
of order $(G |V|)^{-1/2}$.
If the AdS space originates in a configuration of 
a background field (for example the Higgs field), the field fluctuations 
grow until
the energy density becomes infinite and a physical singularity appears.
Notice that both the AdS and AdS-Schwarzschild geometries have 
a timelike boundary, such that the evolution cannot 
be predicted after the bubble wall reaches the boundary, unless additional
conditions are imposed there. 
As a result, a Cauchy horizon appears in the interior of the
bubble. It is expected that,
in the presence of a fluctuating field and beyond the thin-wall approximation, 
a physical spacelike singularity must develop before the Cauchy horizon \cite{freivogel}.  For an AdS-Schwarzschild interior, this spacelike singularity 
is expected to merge with the spacelike singularity at $r=0$. 
We have depicted the merging in fig. \ref{W1}. 
The Penrose diagram supports the conclusion that the endpoint of the evolution
of an expanding bubble
is a catastrophic event. In the context of the Standard Model, such an event 
can be avoided by imposing an upper bound on the scale of inflation 
\cite{tetradis}.

An interesting fact deduced from the second plot of fig. \ref{h1} is
that there exist critical or expanding bubbles with $\ex_M=1$ and
$\zeta >1$. This means that the mass $M$ assigned to the bubble configuration
by an asymptotic observer is smaller than the mass parameter 
of the black hole $m$.

Several other types of Penrose diagrams can be obtained for the other
trajectories depicted in fig. \ref{h1}. They have been analyzed in detail in \cite{tetradis}. It is worth mentioning a particular class of spacetimes
obtained for bubbles with large surface tension $\kx^2>1/\ell_{\rm AdS}^2$ (or $\ex=-1$).
Despite the Newtonian intuition that such bubbles should not expand, general
relativity allows the possibility of a bubble that grows behind 
an event horizon of the external geometry. A faraway observer sees only
a localized defect in spacetime \cite{blau,tetradis}. We do not discuss
such configurations in detail here, as our focus is on bubbles that 
lead to catastrophic events by engulfing the entire external space.

\begin{figure}[!t]
\centering
$$
\includegraphics[width=0.45\textwidth]{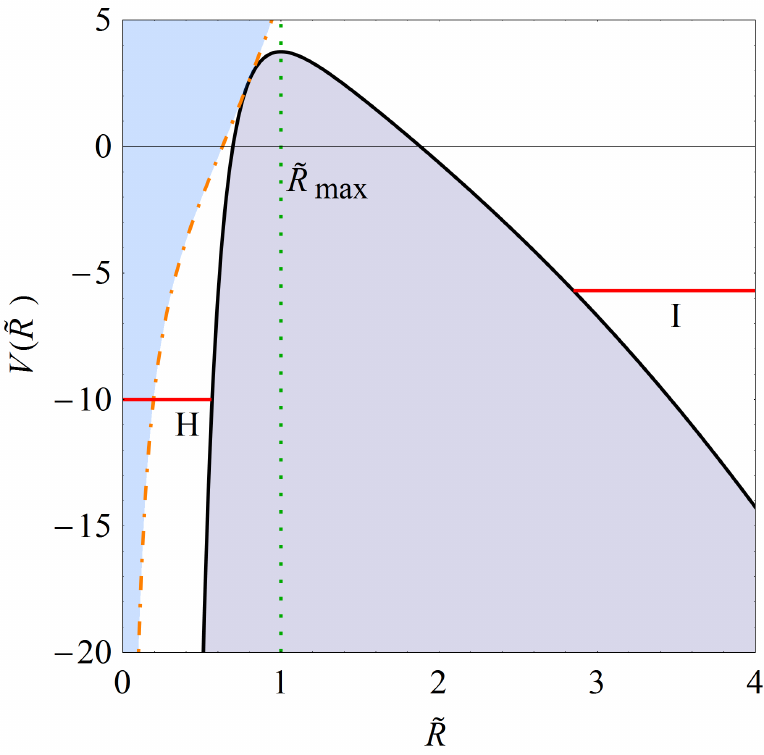}\qquad 
\includegraphics[width=0.45\textwidth]{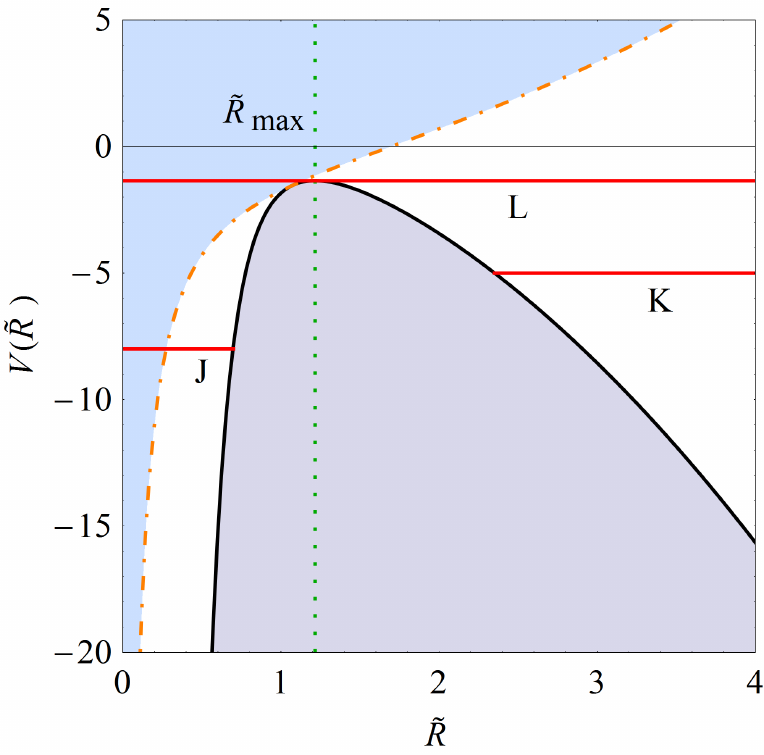}\qquad $$
\caption{\em The `potential' of eq.~(\ref{pot}) for 
$\ex_M=-1$, $\gamma=2$, $\ex=1$, $\delta=0$, $\zeta=0.5$ (left), and
$\ex_M=-1$, $\gamma=2.1$, $\ex=-1$, $\delta=0$, $\zeta=0.5$ (right).}
\label{h12}
\end{figure}

A different class of bubbles are characterized by negative mass $M<0$. Such
solutions are physically acceptable, as the would-be naked singularity at
$r=0$ is eliminated by the interior of the bubble.\footnote{In order to 
avoid naked singularities, we consider only internal geometries with $m>0$.}
Typically, such bubbles have large radii and their energy is dominated by
the negative energy density of the AdS interior. 
Two types of `potentials' for $M<0$ (or $\ex_M=-1$) are depicted in fig. \ref{h12}. The `potential' on the left corresponds to bubbles with relatively
small surface tension ($\ex=1$ or $\kx^2< 1/\ell_{\rm AdS}^2$) and a rather small black
hole at their center ($\zeta=m/|M|=0.5$). The maximum of the `potential' has a
 positive value, which does not allow for solutions of the type C or D of fig. \ref{h1}, because the `energy' $E$ of the solutions is always negative 
 (see eqs. (\ref{pot})). The bubbles either shrink or expand, while there are no critical ones.
The `potential' on the right corresponds to bubbles with large tension 
($\ex=-1$ or $\kx^2> 1/\ell_{\rm AdS}^2$), for which critical bubbles can exist. 
Notice that there is only one event horizon, that of the internal geometry.
 
\begin{figure}[!t]
$$\hspace{-1cm}\includegraphics[width=80mm,height=80mm]{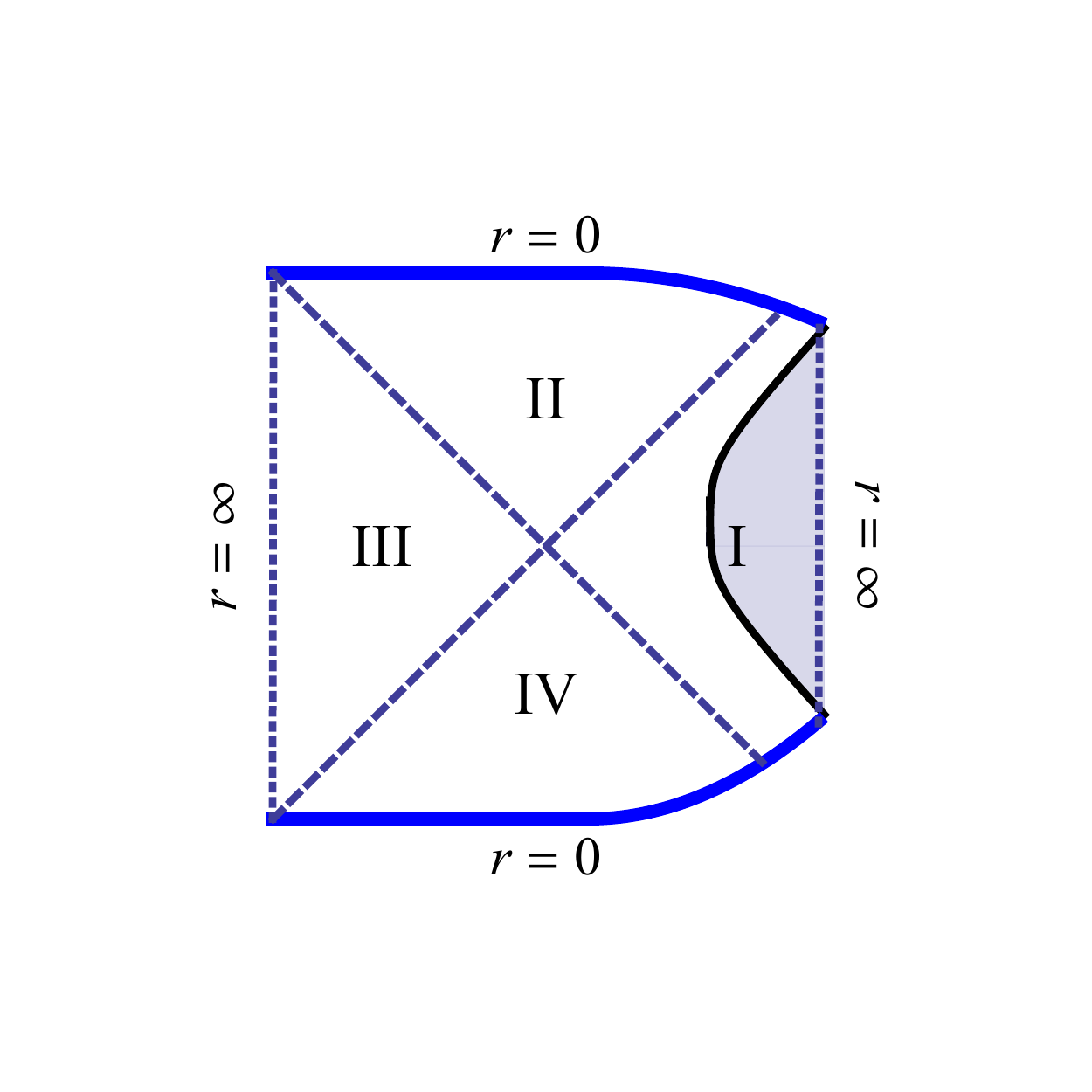}
\includegraphics[width=80mm,height=80mm]{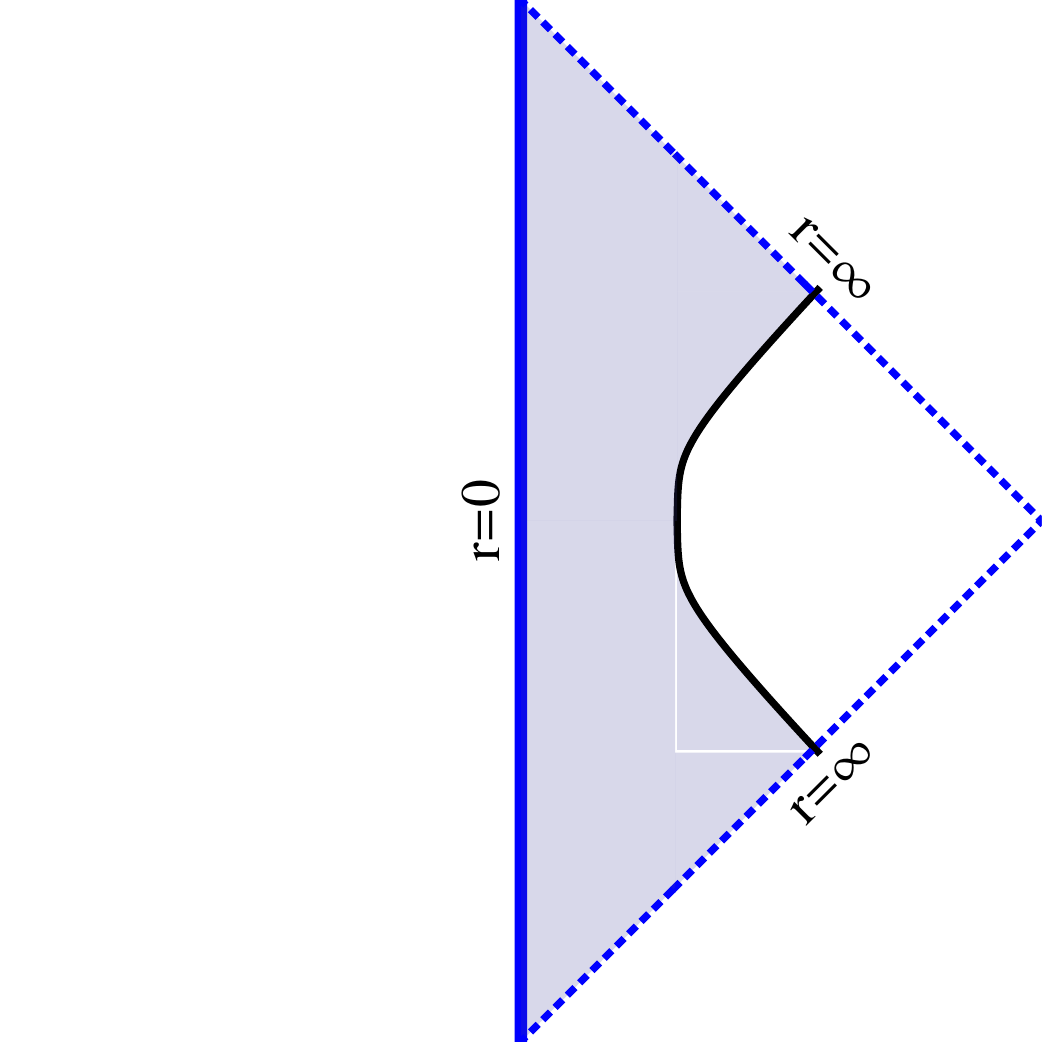} $$
\caption{\em The Penrose diagram for the wall trajectories I and K of fig. \ref{h12}. The total spacetime is constructed by joining the two diagrams, after
the elimination of the shaded areas.  }
\label{W12}
\end{figure}

A typical Penrose diagram, corresponding to the trajectories I and K of
fig. \ref{h12} is depicted in fig. \ref{W12}. The naked singularity in
the diagram on the right is not relevant, as it is eliminated along with the
shaded areas when the global spacetime is constructed.

\begin{figure}[!t]
\centering
$$
\includegraphics[width=0.45\textwidth]{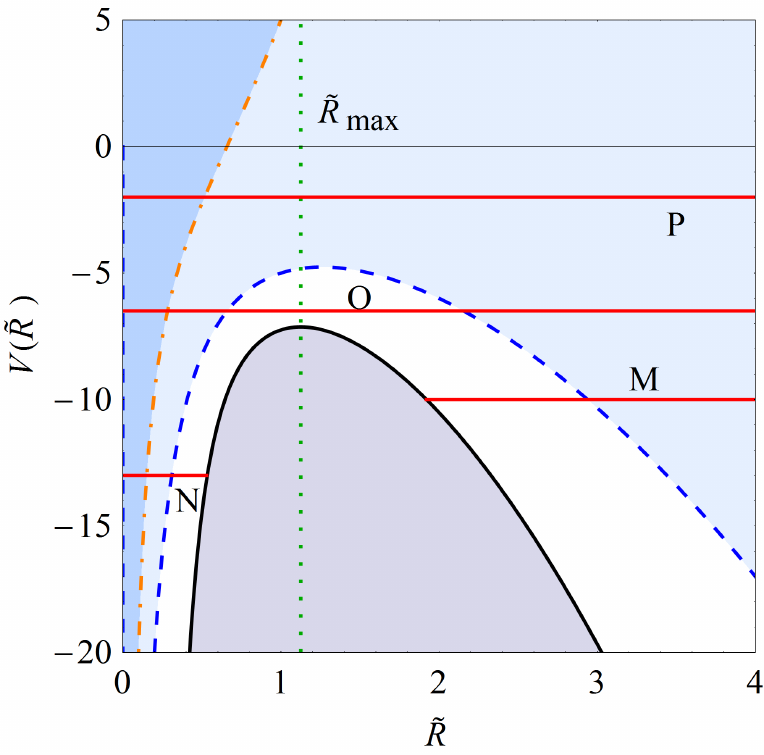}\qquad 
\includegraphics[width=0.45\textwidth]{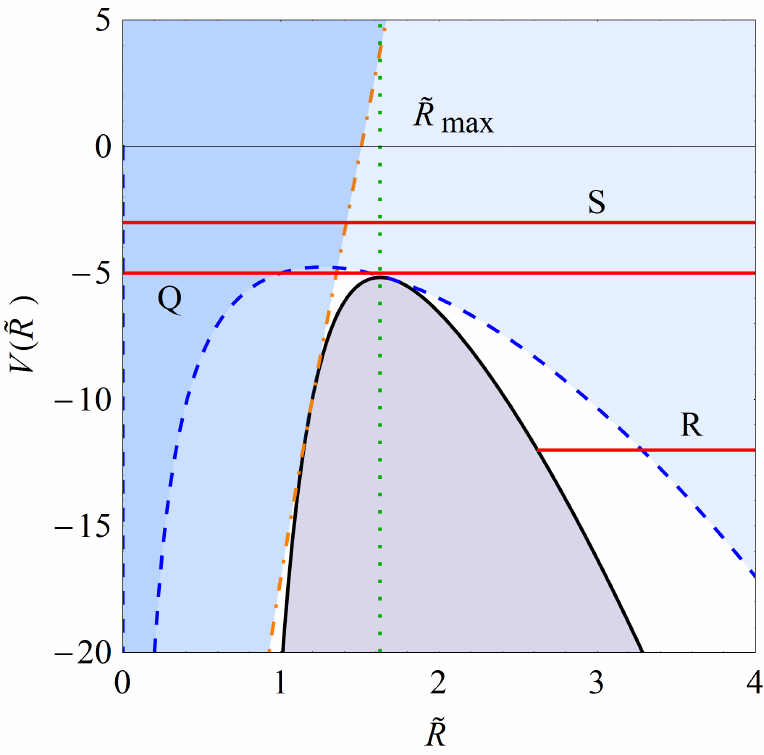}\qquad $$
\caption{\em The `potential' of eq.~(\ref{pot}) for 
$\ex_M=1$, $\gamma=2$, $\ex=1$, $\delta=1$, $\zeta=0.5$ (left), 
and
$\ex_M=1$, $\gamma=2$, $\ex=1$, $\delta=1$, $\zeta=6$ (right).
}
\label{h1aaa}
\end{figure}

\section{Bubble evolution in asymptotically de Sitter space}\label{app3}

We turn next to the case of a bubble evolving in asymptically de Sitter
space, which corresponds to $\delta>0$. 
There are many possibilities depending on the signs of 
$\ex$, $\ex_M$, $\ex_1$, $\ex_2$. We do not perform the complete analysis,
as many cases are similar to those discussed in refs. \cite{blau,tetradis}.
In fig. \ref{h1aaa} we depict the `potential' for some characteristic values of the various parameters for bubble configurations of positive mass $M$. The two
plots correspond to black holes of mass $m$ smaller ($\zeta=m/M=0.5$) or larger ($\zeta=6$) than $M$. We have plotted the `potential' $V(\Rt)$
and indicated the horizon of the internal AdS-Schwarzschild geometry (dot-dashed line), as well as the horizons 
of the external dS-Schwarzschild geometry (dashed line). 
The straight horizontal lines
depict possible configurations of constant 
bubble mass $M$. The line N corresponds to 
a bubble that expands from vanishing radius to a maximal size and then recollapses.
During its evolution the wall crosses the horizon of the  AdS-Schwarzschild geometry and the inner horizon of the dS-Schwarzschild geometry. 
The lines M and R correspond to
bubbles that start with infinite radius, shrink to a minimal size and then
reexpand. Their walls cross only the outer horizon of the dS-Schwarzschild geometry.
The lines O and Q correspond to bubbles whose radii
can vary continuously from vanishing to infinite as a function of time. 
During their evolution they cross the horizon of the internal geometry, as well
as the two horizons of the external geometry. The order in which these 
are crossed 
is different for the two configurations. 
Lines P and S correspond to bubbles whose wall crosses only the horizon of
the AdS-Schwarzschild geometry.
The reason is that `energies' that approach zero correspond to large values 
of the mass parameter $M$. For sufficiently large $M$, the metric function 
$f_{\rm out}(r)$ stays always negative. The 
space has a naked spacelike singularity at $r=0$.  
However, this part of space is eliminated and replaced by the interior of the
AdS bubble.

\begin{figure}[!t]
\centering
$$\hspace{-1cm}\includegraphics[width=80mm,height=70mm]{penrose2.pdf}
\includegraphics[width=100mm,height=60mm]{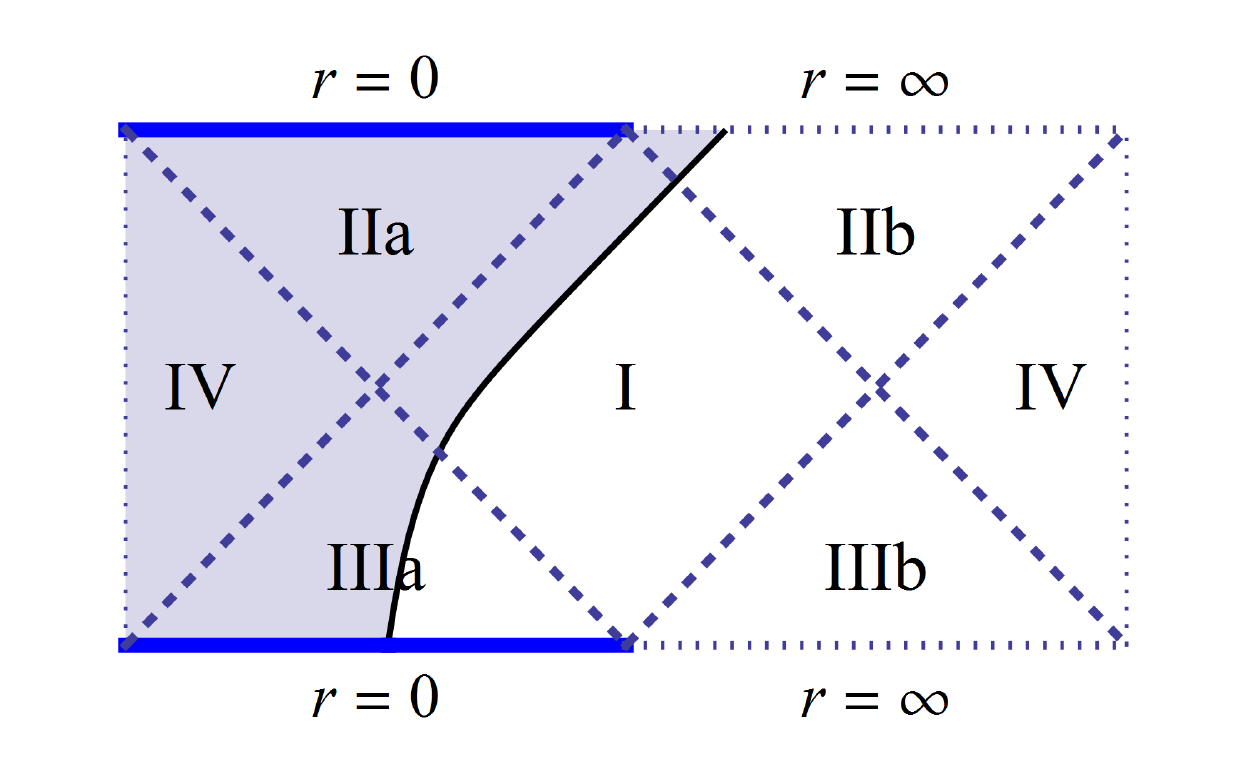} $$
\caption{\em The Penrose diagram for the wall trajectory O of fig. \ref{h1aaa}. The total spacetime is constructed by joining the two diagrams, after
the elimination of the shaded areas.  }
\label{hh12}
\end{figure}

In fig. \ref{hh12} we present the Penrose diagram for the trajectory O 
of fig. \ref{h1aaa}.
Thick blue lines denote curvature singularities, 
the dashed lines horizons and the dotted lines conformal infinities.
The two thin vertical lines at the ends of the right
diagram indicate that the pattern is repeated indefinitely on either side. 
A similar diagram corresponds to the trajectory Q. 
The wall of the expanding bubble crosses all three horizons and the bubble
reaches infinite radius. 
The question whether the AdS bubbles can completely eliminate the
surrounding dS space was addressed in ref. \cite{tetradis} in the
absence of a central black hole. The conclusion is
the same for an internal AdS-Schwarzschild geometry:
It is apparent from fig.~\ref{hh12} that asymptotically the wall trajectory reaches
spacelike infinity. The wall location separates two spacelike regions: the first 
corresponds to the interior of the bubble, while the second is 
part of an asymptotically dS spacetime.  
The total spacetime 
contains large AdS bubbles within large 
dS regions. In other words, 
expanding bubbles are inflated away: they expand, but the dS space between them also grows. This scenario is in contrast to the case of 
asymptotically flat spacetime, in which the wall asymptotically
reaches null infinity and the whole space is
engulfed by the AdS bubbles.

Another important question concerns 
the consequences for an outside observer of the AdS `crunch' in the bubble interior. This issue was also addressed in detail in ref. \cite{tetradis}.
It was shown that the singularity never 
reaches the wall, as the latter expands with the speed of light. 
From the point of view of an external observer, the bubble just expands forever,
within either de Sitter or Minkowski spacetime.

    \end{document}